\documentclass[a4paper,twocolumn,11pt]{quantumarticle}
\pdfoutput=1
\usepackage[utf8]{inputenc}
\usepackage[english]{babel}
\usepackage[T1]{fontenc}
\usepackage{amsmath, amssymb}
\DeclareMathOperator*{\argmax}{arg\,max}

\usepackage[caption = false]{subfig}
\usepackage{tikz}
\usetikzlibrary{quantikz2}
\usepackage{physics}
\usepackage{algorithm2e}
\usepackage[numbers,sort&compress]{natbib}
\usepackage{hyperref}

\usepackage{tikz}
\usepackage{lipsum}

\begin{document}
\RestyleAlgo{ruled}

\title{Gate Sequence Optimization for Parameterized Quantum Circuits using Reinforcement Learning}
\author{Tom~R.~Rieckmann}
\affiliation{Institute for Physics, University of Rostock, Albert-Einstein-Straße 23-24, 18059 
Rostock, Germany}
\email{tom.rieckmann@uni-rostock.de}
\author{Stefan~Scheel}
\affiliation{Institute for Physics, University of Rostock, Albert-Einstein-Straße 23-24, 18059 
Rostock, Germany}
\author{A.~Douglas~K.~Plato}
\affiliation{Institute for Physics, University of Rostock, Albert-Einstein-Straße 23-24, 18059 
Rostock, Germany}

\begin{abstract}
    Current experimental quantum computing devices are limited by noise, mainly originating from 
    entangling gates. If an efficient gate sequence for an operation is unknown, one often employs 
    layered parameterized quantum circuits, especially hardware-efficient ans\"atze, with fixed 
    entangling layer structures. 
    We demonstrate a reinforcement learning algorithm to improve on these by optimizing the 
    entangling gate sequence in the task of quantum state preparation. This allows us to restrict the 
    required number of CNOT gates while taking the qubit connectivity architecture into account. 
    Recent advancements using reinforcement learning have already demonstrated the power of this 
    technique when optimizing the circuit for a sequence of non-parameterized gates. We extend this
    approach to parameterized gate sets by incorporating general single-qubit unitaries, thus 
    allowing us to consistently reach higher state preparation fidelities at the same number of CNOT 
    gates compared to a hardware-efficient ansatz.
    
\end{abstract}

\section{Introduction}\label{Introduction}
Quantum computing has made major advancements in recent years, with multiple systems claiming to 
have reached quantum primacy \cite{Zhu2022, Madsen2022, Morvan2024}.
However, these results have been achieved on artificial problems such as random circuit sampling 
\cite{Boixo2018} without any known 'real world' applications. Current systems still fall short of 
being able to perform the tasks at the heart of quantum computing development such as Shor's 
algorithm \cite{Shor1994, Monz2016}, quantum simulation \cite{Tacchino2020}, and quantum 
machine learning \cite{PeralGarcia2024}.
While the recent experimental demonstrations of error correction are promising 
\cite{Acharya2024}, a full scale implementation is still out of reach for today's noisy intermediate-scale quantum (NISQ) systems \cite{Preskill2018}.
\par Consequently, there is a strong interest in finding algorithms that can yield useful 
results without the need for fault tolerance through error correction \cite{Preskill2018, Lanes2025}. Recent work on large interacting systems, where classical approaches struggle, have demonstrated methods for calculating expectation values \cite{Kim2023} and ground state energies \cite{RobledoMoreno2025}. However, noise is still a major limiting factor for the maximally viable scale of the calculations and one requires error mitigation techniques to get meaningful results on current devices. These techniques, such as zero noise extrapolation or readout error mitigation, rely on postprocessing of repeated circuit evaluations to improve the directly measured results \cite{Cai2023}. 
Instead, one of the most obvious paths to avoiding noise is to optimize the algorithm's quantum circuit. This requires defining a useful optimization target. 
Our main goal is to minimize the number of entangling gates, as for most experimental systems these are the dominant source of errors \cite{Popkin2016}. We expect that, as a by-product, the depth of the circuit is also kept small, which is particularly important for systems with short dephasing and relaxation times compared to the gate execution time, such as is the case with superconducting qubits.

Optimization of circuits is a well-known task with tools provided by many quantum 
computing frameworks \cite{JavadiAbhari2024, Hua2023}. However, such methods can be computationally 
expensive, with recent work suggesting that most quantum circuit optimization problems are 
NP-hard \cite{Wetering2024, Wetering2025optimalcompilation}. 
Finding efficient implementations is particularly difficult when the exact unitary to be implemented is not yet fully determined by a gate sequence, as is the case for variational quantum 
algorithms \cite{Cerezo2021}, such as variational quantum eigensolvers \cite{Kandala2017, Tilly2022} and quantum machine learning \cite{Benedetti2019, Abbas2021}. 
One approach is to make use of parameterized quantum circuits, where the general structure is fixed, but variable gates, such as angle-dependent rotations, allow for control over the output distributions. This relies on finding an ansatz with sufficient expressibility for all desired operations. \par 

A variety of gate sequences have been proposed, which can implement any unitary or state by varying only their continuous parameters, and lower bounds for the minimally required number of entangling gates for this task have been determined \cite{Barenco1995, Bergholm2005, Plesch2011, Zhang2022}.
More recently, circuits created by repeating layers of local gates and layers of a fixed entangling gate sequence to approximate any desired operation have become popular. They can significantly reduce the gate number, while still being highly 
expressive in the types of operations they can represent \cite{Sim2019, Du2020}.
Even though this one-size-fits-all approach is very simple to use and yields good results, 
it is by no means optimal, particularly when considering the properties and limitations of individual experimental systems. \par 
Instead, we propose a reinforcement learning (RL) algorithm to optimize the entangling gate sequence for a parameterized universal gate set, and apply it to the problem of quantum state preparation. 
RL has previously been applied to optimizing circuits, initially limited to the
Clifford gate set \cite{Fuerrutter2024, Kremer2024}, as well as unitary synthesis \cite{Rietsch2024}. Current research has taken a focus on the discrete Clifford+T gate set \cite{Riu2025reinforcement, Ruiz2025}. These works have demonstrated that such 
algorithms can generalize to the entire problem space from only a limited number of training 
simulations while taking CNOT connectivity architectures into account. Our work differs by including arbitrary parameterized single-qubit gates, which makes our gate set more similar to the native gate set employed on many experimental systems, such as those developed by IBM \cite{JavadiAbhari2024}. This allows us to investigate the effectiveness of RL for optimizing circuits used in variational quantum algorithms.
\par 
In this paper, we will first introduce the theoretical background and experimental limitations for 
quantum state preparation. We then detail the design of the RL agent.
Results are compared to layered parameterized circuits with common hardware-efficient entangling 
sequences. In our analysis we have placed the focus on the properties of publicly available quantum computers from IBM, but the general approach can be deployed on any system using a gate-based computation model.

\section{Optimal Quantum State Preparation}
In quantum state preparation, the goal is to achieve a target $\ket{\psi_T}$ starting from the 
initial state of the experimental system (usually the n-qubit computational basis 
state $\ket{0}$).
With a universal quantum gate set, an exact solution for this task is 
guaranteed to exist when neglecting noise \cite{Kitaev1997, Nielsen2010}.  \par 
On NISQ devices noise cannot be ignored, thus any state preparation will not be exact. To quantify 
the accuracy we use the fidelity 
\begin{equation}
F(\rho, \sigma) = \left(\tr(\sqrt{\rho^{1/2} \sigma \rho^{1/2} })\right)^2,
\end{equation}
which simplifies to $F(\rho, \ket{\psi}) = \bra{\psi}\rho\ket{\psi}$ if one of the states is pure 
\cite{Nielsen2010}. \par 
While entangling gates, such as the CNOT gate, generally have the biggest impact on fidelity, 
additional care has to be taken when constructing the circuit to consider the architecture of the 
specific device. Due to experimental limitations, one cannot choose any pair of qubits as control 
and target.

\begin{figure}
   \centering
   \includegraphics[width=1\linewidth]{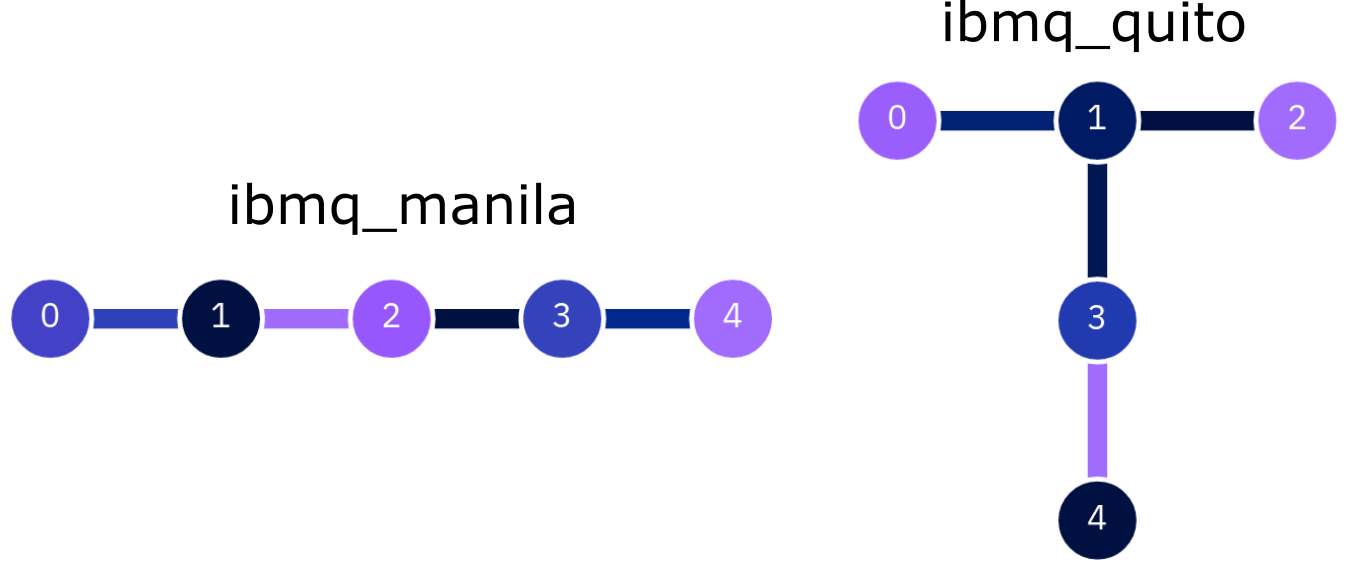}
   \caption{CNOT connectivity for ibm\_manila and ibm\_quito. Numbers represent qubits, connections depict CNOTs connections in the hardware. The colour coding is given by gate errors (connections) and decoherence (qubits). Images taken from IBM Quantum \cite{JavadiAbhari2024}.}
   \label{CNOTConnectivity}
\end{figure}

\begin{figure*}
\centering
\subfloat[]{%
    \begin{tikzpicture}
    \node[scale=1.2] {
    \begin{quantikz}[row sep={1cm,between origins}]
     &  \gate{U} & \gate[4, disable auto height]{U_{\text{layer}}^L} & \\
     &  \gate{U} &                              & \\ 
     &  \gate{U} &                              & \\ 
     &  \gate{U} &                              & 
    \end{quantikz}};
    \end{tikzpicture}}%
\hspace{0.5cm}\subfloat[]{%
    \begin{tikzpicture}
    \node[scale=1.2] {
    \begin{quantikz}[row sep={1cm,between origins}]
     & \ctrl{1} &  & & \gate{U} & \\
     & \targ{} & \ctrl{1} & & \gate{U} & \\
     & & \targ{} & \ctrl{1} &  \gate{U} & \\
     & & & \targ{} & \gate{U} & 
    \end{quantikz}};
    \end{tikzpicture}}%
\hspace{0.5cm}\subfloat[]{%
    \begin{tikzpicture}
    \node[scale=1.2] {
    \begin{quantikz}[row sep={1cm,between origins}]
     &  \ctrl{1}    &            &              & \gate{U} & \\
     &  \targ{}     &   \gate{U} &  \ctrl{1}    & \gate{U} & \\
     &  \ctrl{1}    &   \gate{U} &  \targ{}     & \gate{U} & \\
     &  \targ{}     &            &              & \gate{U} &
    \end{quantikz}};
    \end{tikzpicture}}%
\caption{Makeup of a layered parameterized circuit. Single-qubit gates $U$ are parameterized with up to three independent parameters each. (a) depicts the general structure, which consists of layers $U_{\text{layer}}$ repeated $L$ times. As layers begin with entangling gates, a local rotation layer is added at the beginning. In (b) (linear layer) and (c) (pairwise layer) one can see two popular variants for $U_{\text{layer}}$, which will be used for the performance comparison of our algorithm. }\label{FigLayeredExamples}
\end{figure*}
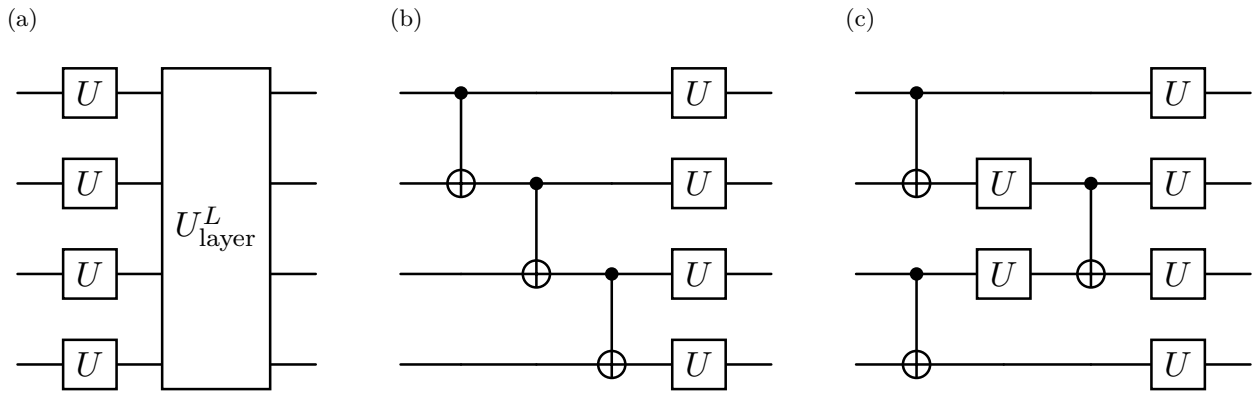

\subsection{Gate Set Limitations}
Entangling gates that are directly implementable on hardware are determined by the system's qubit 
connectivity architecture. CNOT gate connectivity examples for two previous IBM Quantum systems can be seen in Figure \ref{CNOTConnectivity}.
For the remainder of this paper we will restrict entanglement operations to CNOT gates, though all statements regarding its properties and usage can also be applied to equivalent\footnote{We consider the entangling gates equivalent if one can be derived from the other by only adding single qubit rotations before and after. Our algorithm can be used identically by simple replacing the CNOT gates after the fact and simplifying neighbouring single qubit rotations.} entangling gates such as the ECR gate \cite{JavadiAbhari2024}.

Circuits that make use of gates not directly available on the device have to be mapped to the 
architecture. The most direct approach is to use SWAP gates, which each contain up to 3 CNOTs. While more sophisticated approaches, such as those implemented in the IBM transpiler 
\cite{JavadiAbhari2024}, may yield much better results, these mapping procedures should be avoided 
during circuit design. This becomes even more important for larger circuits with respect to depth 
and width, as the likelihood for an efficient relabelling of qubits decreases. \par

\subsection{Layered Parameterized Circuits}
Hardware-efficient ans\"atze are designed to avoid these transpiling issues by densely stacking 
CNOT gates in a nearest neighbour architecture. This forms an entangling 
layer $U_{\text{ent}}$, which is interleaved with local rotation layers $U_{\text{rot}}$. In \cite{Sim2019} a variety of different entangling layers were compared in terms of their expressibility and entangling capability. Examples for two of the most popular structures can be seen in Figure \ref{FigLayeredExamples}.
By stacking enough layers $U_{\text{layer}} = U_{\text{rot}}(\vb*{\theta}) U_{\text{ent}}$, universality of an ansatz is guaranteed. A unitary is implemented by 
\begin{equation}
    U =  \left(\prod_l^L  U_{\text{rot}} (\vb*{\theta}_l) U_{\text{ent}}\right) U_{\text{rot}} (\vb*{\theta}_0),
\end{equation}
with 
\begin{equation}
    U_{\text{rot}} = U_{\text{local}}^{\otimes n}.
\end{equation} 
Often 
\begin{equation}\label{Eq:EfficientSU2Local}
    U_{\text{local}}(\theta_1, \theta_2) = R_z(\theta_1) R_y(\theta_2)
\end{equation} 
is sufficient instead of an arbitrary single-qubit rotations, as Eq. (\ref{Eq:EfficientSU2Local}) can reach any possible output provided that one does not start in an eigenstate of $R_y$. 
The free parameters of each local rotation layer $\vb*{\theta}_l$ are determined using classical optimization methods. Optimization techniques designed for parameterized quantum circuits, such as quantum natural gradient, may converge in fewer steps than standard gradient descent or ADAM \cite{Stokes2020}. 

\section{Method}
\begin{algorithm}[t]
	\DontPrintSemicolon
	\caption{Circuit Generation}\label{AlgCircGen}
	\KwData{Target $\rho_T$, empty circuit $C$, agent (NN, outputs $Q_W$), state loss $L$, maximum number of actions $M$}
	\KwResult{Populated circuit $C$}
	Add local rotation layer $U_{\text{rot}}(\vb*{\theta}_0)$ to $C$\;
	Minimize $L(U_{\text{rot}}\rho_T U_{\text{rot}}^\dag)$ to determine $\vb*{\theta}_0$\;
	Find $\rho_1 = U_{\text{rot}}\rho_T U_{\text{rot}}^\dag$\;
	\For{$k = 1, 2, \ldots M$}{
		Observe input $\vb{v}_{\text{in}}$ from $\rho_k$ of $s_k$ \;
		Get state-action values $\vb{Q}_W(s_k, a)$ from NN using $\vb{v}_{\text{in}}$ as input \;
		Sample $a_k$ from $\vb{Q}_W(s_k, a)$\;
		\If{$a_k$ is stop}{
			Set $N=k-1$\;
			break}
		Get $U_k$ from $U_{\text{CX}}$ of chosen CNOT and $U_{\text{local}}$ on control, target qubits\;
		Minimize $L(U_{k}(\vb*{\theta}_k)\rho_k U_{k}^\dag(\vb*{\theta}_k))$ over $\vb*{\theta}_k$\;
		Set $\rho_{k+1} = U_k\rho_k U_k^\dag$\
	}
	Full Circuit $C$ defines $U_C$\;
	Minimize $L(U_C\rho_T U_C^\dag)$ to redetermine all $\vb*{\theta}_0,\vb*{\theta}_1, \ldots \vb*{\theta}_N$ in $U_C(\vb*{\theta}_0,\vb*{\theta}_1, \ldots \vb*{\theta}_N)$\;
	Add NOT gates to go from closest computational basis state $\ket{b}$ to $\ket{0}$\;
	Invert circuit for state preparation task with $\rho_{\text{final}} = U_C^\dag \dyad{0}{0}U_C$
\end{algorithm}

While the layered parameterized circuit approach has proven successful, it requires significant redundancy in the number of gate operations in order to express a wide range of quantum states. In addition, the choice of layer structure is also limited by the available hardware architecture. Both these issues can be overcome by employing an adaptive circuit design strategy, where an algorithm first determines an optimal gate structure, before standard optimisation techniques are used to fix the continuous gate parameters. This problem is particularly suited to reinforcement learning (RL), as it does not require a priori representative datasets of known efficient quantum circuits – which are computationally expensive to generate. Our approach is based on a Double Deep Q-Network (DDQN) agent \cite{Hasselt2016}, which builds circuits sequentially by selecting two-qubit operations (actions) using an evaluation by a neural network (NN). At every step, $k$, in the circuit generation process, we can choose to either stop (if the circuit is deemed to be complete), or apply a parameterizable two-qubit gate drawn from a predetermined allowable gate set. Our goal is achieving an arbitrary target state with as few entanglement operations as possible. For practical reasons we consider the inverse problem of transforming a target state back to the ground state, as this simplifies the optimization and evaluation. The objective then becomes finding a policy, which at each step chooses the action which is most likely to lead to the ground state in the least number of overall steps. Note, this need not be the action with the best immediate improvement. \par 
The process of circuit generation from a trained NN is explained in Section \ref{SubSecCircuitGen}, with a summary given by Algorithm \ref{AlgCircGen}. Section \ref{SubSecRLAgent} provides a detailed description of our agent. Training parameters can be found in \cite{Rieckmann_Gate_Sequence_Optimization_2025} or by reaching out to the authors.

\subsection{Circuit Generation}\label{SubSecCircuitGen}

The neural network evaluation requires a (real) vector input containing information about the current state of the system, $s_k$. As we are working with the inverse problem, we do not need to provide $\rho_T$ at each step, and instead only keep track of the current density matrix, $\rho_k$, along with an optional list of previously applied actions, $\{a_j\}_{j<k}$. Thus, we construct the NN input at step $k$ as, 
\begin{equation}\label{EqInput}
    \vb{v}_{\text{in}}(s_k) = \pmqty{\text{vec}(\rho_k - \frac{\mathbb{I}}{d})\\
                                     c_{\text{in}}\text{key}(a_0) \\
                                     \vdots \\
                                     c_{\text{in}}\text{key}(a_{k-1}) \\
                                     0 \\
                                     \vdots}, 
\end{equation}
where the vectorisation is over the real and imaginary parts of $\rho_k-\frac{\mathbb{I}}{d}$,  $d$ is the dimension of the Hilbert space and $\text{key}(a)$ are zero mean vectors uniquely identifying each possible action, constructed from a binary representation of the NN output vector index associated to that action. The subtraction of a trace one matrix from $\rho_k$, together with the normalisation of the action keys, ensures that, in accordance with standard data preprocessing procedures \cite{Bishop2023}, each feature has an expected mean of zero over datasets of Haar random states. We also introduce a scaling factor $c_{\text{in}}$, set during training, which determines the relative importance given to past actions vs. the current density matrix (its role will be described in more detail in Section \ref{SubSecRLAgent}). Finally, to maintain constant input dimensions, the input vector is padded with zeros corresponding to future actions.

The objective of the algorithm is to find a gate sequence which diagonalises the target density matrix in the computational basis. From this one can readily reach the ground state, $\ket{0}$, by appropriate application of Pauli $X$ operations to relevant qubits. Thus, a suitable loss function for determining gate parameters is given by, 
\begin{equation}\label{BFGSLoss}
    L(\rho) = \sum_i \sum_{j = i+1} \abs{\rho_{ij}}^2,
\end{equation}
which can be minimised using the Broyden–Fletcher–Goldfarb–Shanno (BFGS) algorithm. The initial data for the NN input is determined by first applying a layer of parameterized local unitaries, $U_{\text{rot}}$, to the $n$ qubit target state $\rho_T$, and then minimising $L(U_{\text{rot}}\rho_T U_{\text{rot}}^\dag)$ to obtain $\rho_1$. This provides $\vb{v}_{\text{in}}(s_1)$. The sequence of entangling gates is then determined by repeatedly applying unitaries corresponding to actions $a_k$ sampled from our RL agent, updating $\vb{v}_{\text{in}}$ at each step. The sampling process makes use of the NN output vector, which is trained (see Section \ref{SubSecRLAgent}) to provide an estimate of the exact state-action values $Q_{\text{true}}$ defined by,
\begin{align}
	Q_{\text{true}}(s_k,a_k) & = r_k + \gamma V_{\text{true}}(s_{k+1}), 
\end{align}
where $r_k$ is the reward signal. $V_{\text{true}}$ represents the value/desirability of the next state and $\gamma<1$ is the discount factor, which can be tuned to make the agent prioritize short episodes. In the final state $s_N$ of an episode, the state-action value is $Q_{\text{true}}(s_k,a_k) = r_k$.

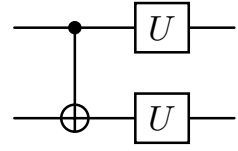
\begin{figure}
	\centering
	\begin{tikzpicture}
		\node[scale=1.2] {
			\begin{quantikz}[row sep={1cm,between origins}]
				& \ctrl{1} & \gate{U} & \\
				& \targ{} & \gate{U} &
		\end{quantikz}};
	\end{tikzpicture}
	\caption{Individual action which we use for our results. Local gates are only added on target and control qubits of an entangling gate. Other single-qubit operations or entangling gates may be chosen as elementary building blocks, such as the Toffoli gate or the controlled-Z gate.}
	\label{FigIndividAction}
\end{figure}

The optimal strategy maximizing future rewards is the one that always picks the action $a_k$ with the highest "true" state-action value. Under this policy $V_{\text{true}}(s_{k+1}) = \max_{a}  \vb{Q}_{\text{true}}(s_{k+1}, a)$, i.e.
\begin{align}
	Q_{\text{true}}(s_k,a_k) & = r_k + \gamma \max_{a}  \vb{Q}_{\text{true}}(s_{k+1}, a) \\
    & = \sum_{l=k}^N \gamma^{l-k} r_l.
\end{align}
Note that $\vb{Q}(s_k, a)$ represents a $d_{\text{out}}$-dimensional vector containing the values of all actions in state $s_k$, and $a_k$ represents a specific action associated to an entry of this vector. Thus $Q(s_k,a_k)$ and $\max_{a}\vb{Q}(s_{k+1}, a)$ are scalar. For a given architecture, we set $d_{\text{out}}$ to the total number of allowable CNOT gates plus one (for the stopping action), though in principle additional gate combinations could be included. Thus each action $a_k$ is mapped to a unitary $U_{k} \equiv U_{a_k}$ composed of a CNOT gate, $U_{\text{CX}}$, which we then append with local parameterized unitaries on the control and target qubits in order to guarantee universality (Fig. \ref{FigIndividAction}). The set of all continuous gate parameters for the action are collectively labelled $\vb*{\theta}_k$ and determined at each step by minimizing the loss, Eq. (\ref{BFGSLoss}), using the BFGS algorithm, which yields the next $\rho_{k+1} = U_k (\vb*{\theta}_k) \rho_k U_k^\dag (\vb*{\theta}_k)$. \par

\begin{figure*}
  \centering
	\includegraphics[width=0.8\linewidth]{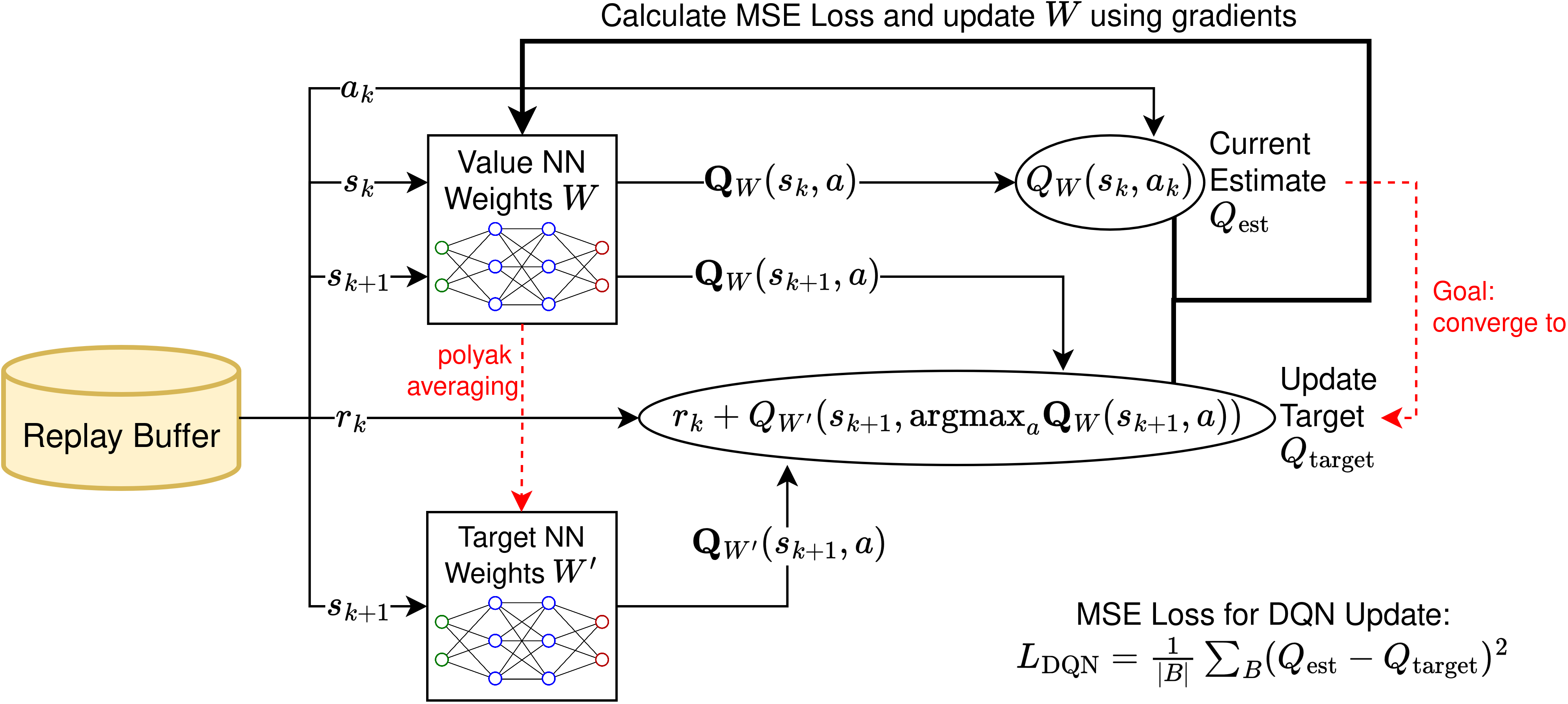}
	\caption{Schematic depicting the process of updating the NN weights from data stored in the replay buffer. The procedure is not performed for a single set of $(s_k, a_k, r_k, s_{k+1})$ but rather for a batch $B$. The target NN is obtained from the value NN using polyak averaging, i.e. after an update step the new target NN weights are calculated as a weighted average of its (99\%) and the value NN's (1\%) weights.}
 	\label{FigDDQNSchematic}
\end{figure*} 

With only such local optimizations, the generated circuits are typically inefficient in terms of their gate number, as the optimal gate parameters are dependent on the future gate sequence. 
To account for this, the final state is determined by performing a global minimization of (\ref{BFGSLoss}) over all continuous parameters, once the complete gate sequence has been established. A comparison of the performance with and without global optimization is given in Table \ref{tab:SimRealComparison}. 
Due to its high computational cost, we perform this minimization only once after the entire gate sequence has been determined. The state preparation can then easily be performed using the inverse circuit, which should yield states as accurate to $\rho_T$ as possible. By evaluating the circuit with a reward $r_k$ proportional to the fidelity between the final state $\rho_{\text{final}}$ and the target $\rho_T$ combined with a discount $\gamma<1$, we can create an environment where the agent is not only focused on maximizing the fidelity, but also on finding the shortest gate sequence to do so.\par
Finally, to guarantee the desired accuracy, we add a fidelity threshold $T_F$, which must be reached in order to receive a reward
\begin{equation}\label{RewardFunc}
    r_{k}(s_k) = \begin{cases}
                    c_r F(\rho_{\text{final}}, \rho_{T})& \text{if } k=N \\ & \text{\& } 1-F<T_F \\
                    0  &  \text{else.}
                \end{cases}
\end{equation}
Actions during the episode receive no reward or penalty.
$c_r$ is a simple scaling factor. In practice, the final step reward $r_N$ can more easily be calculated by
\begin{equation}
    F(\rho_{\text{final}}, \rho_{T}) = \ev{U_C\rho_T U_C^\dag}{0}
\end{equation}
due to the invariance of the fidelity under unitary transformations.

\subsection{Agent Structure}\label{SubSecRLAgent}
The remaining task is to train the weights $W$ of the NN, so that its output vector, used as the estimate $Q_W(s_k, a_k)$, is very close to $Q_{\text{true}}$.
Training data samples are generated by performing the circuit generation for Haar-random target 
states. State sampling includes partially separable states as well, e.g. only $m$ of all qubits 
might be entangled by a Haar-random $m$-qubit unitary, whereas the other qubits are in random single-qubit states. To avoid bias, the positions of qubits are then shuffled randomly. We utilise 
the $\varepsilon$-greedy algorithm for action sampling to facilitate exploration, i.e. with probability $\varepsilon$ we choose random actions, otherwise the highest value (the greedy action) is selected. The vectors representing states $s_k$ and 
$s_{k+1}$, actions $a_k$, and rewards $r_k$ are stored in a replay buffer. \par 
When updating the network, a batch of individual steps from all stored episodes is sampled from the replay buffer with uniform probability. This prevents the agent from focusing only on recently sampled training data. A simplified schematic depicting the remaining steps of the update procedure of the DDQN agent is given in Figure \ref{FigDDQNSchematic}. 
As one can see, the sampled data is used to calculate the update targets
\begin{multline}
	Q_{\text{target}}(s_k,a_k) = r_k +  \gamma Q_{W'}(s_{k+1}, \\\argmax_a \vb{Q}_W(s_{k+1}, a))
\end{multline}
where $Q_{W'}$ represents the output of the target NN with weight $W'$, which is a delayed copy of the action selection NN and is obtained by polyak averaging of $W'$ and $W$. This improves training stability by smoothing out rapid changes in the state-action value targets $Q_{W}$. NN weights $W$ are updated using Adam gradient descent on the loss
\begin{equation}\label{EqMSELoss}
    L_{\text{DQN}} = \frac{1}{\abs{B}}\sum_B (Q_W(s,a)-Q_{\text{target}}(s,a))^2
\end{equation}
of a batch $B$.

On top of these standard protocols we have used the following procedures to improve training 
efficiency. Firstly, we added the possibility of prioritized exploration to the standard 
$\varepsilon$-greedy algorithm.
In many states, some gates perform significantly worse than others, e.g. when an individual qubit is already separable. By adding prioritized explorations we make it 
less likely that these actions are sampled, but still keep a high exploration rate of interesting 
states, e.g. where the agent cannot distinguish between the efficacy of only two actions. Our 
implementation adds a chance to randomly sample from only the actions with the highest state-action value estimates. \par
Secondly, instead of starting with a low threshold $T_F$ as dictated by our desired accuracy, we 
always use a high threshold to begin with and decrease it by exponential decay during training, if the agent is successful enough on the previous threshold. This is necessary, as when we train from scratch it is difficult to satisfy a high threshold by randomly guessing sequences. \par 

\begin{figure}
	\centering
	\includegraphics[width=1\linewidth]{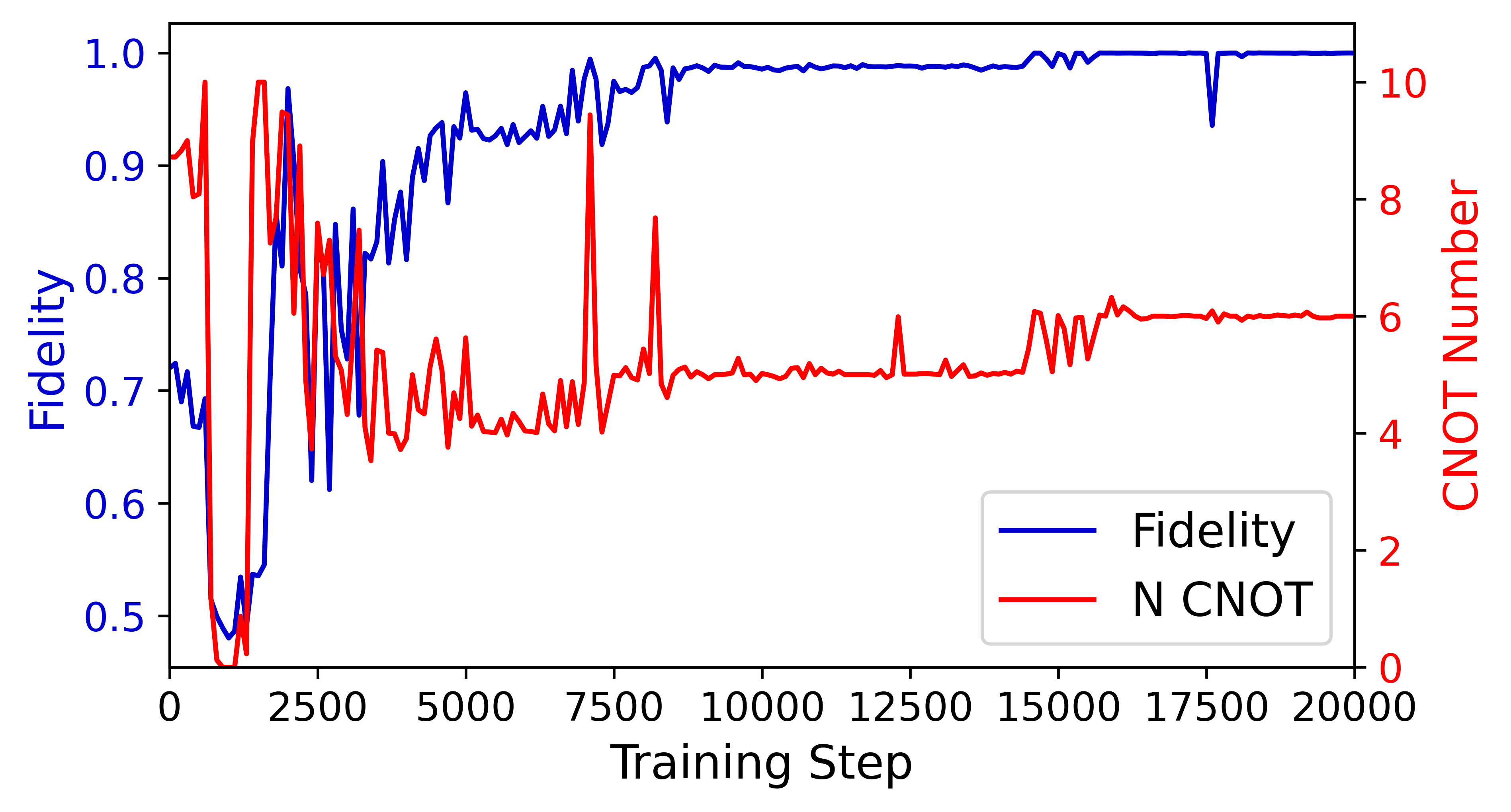}
	\caption{Performance on fully entangled Haar-random states over the course of training for a four-qubit system with unrestricted CNOT connectivity, i.e. all possible CNOT gates are available. The agent slowly improves fidelity by adding more CNOT gates, should they become necessary.}
	\label{FigSampleTraining}
\end{figure}

Moreover, we have observed that adding information of previous actions to the input may result in the agent focusing too much on the action sequence and not considering the quantum state. By reducing the factor $c_{\text{in}}$ from 
Eq. (\ref{EqInput}) in large infrequent steps after meeting the targeted threshold $T_F$, the agent is forced to focus more on $\rho_k$, leading to an adaptation of the gate sequence for individual target states.

\section{Results}
Training has been performed for systems with up to five qubits. Though only select architectures were tested, we expect the algorithm should converge to solutions for any, arbitrarily chosen, universal connectivity. 
Performance was evaluated by running the trained agents on a separate batch of 1000 Haar-random 
states. For this evaluation, we forgo exploration and have agents select greedy actions only, i.e. we set $\varepsilon=0$. An 
example of the mean fidelity and number of CNOT gates used by the agent over the course of training can be seen in Figure 
\ref{FigSampleTraining}. The latter is the main 
factor determining expressivity, thus more CNOT gates typically result in higher fidelities to the 
target state. Agents add gates if they are needed to match the threshold, but due to the penalty from the discount 
factor, unnecessary gates are avoided.

\begin{figure}
	\centering
	\includegraphics[width=1\linewidth]{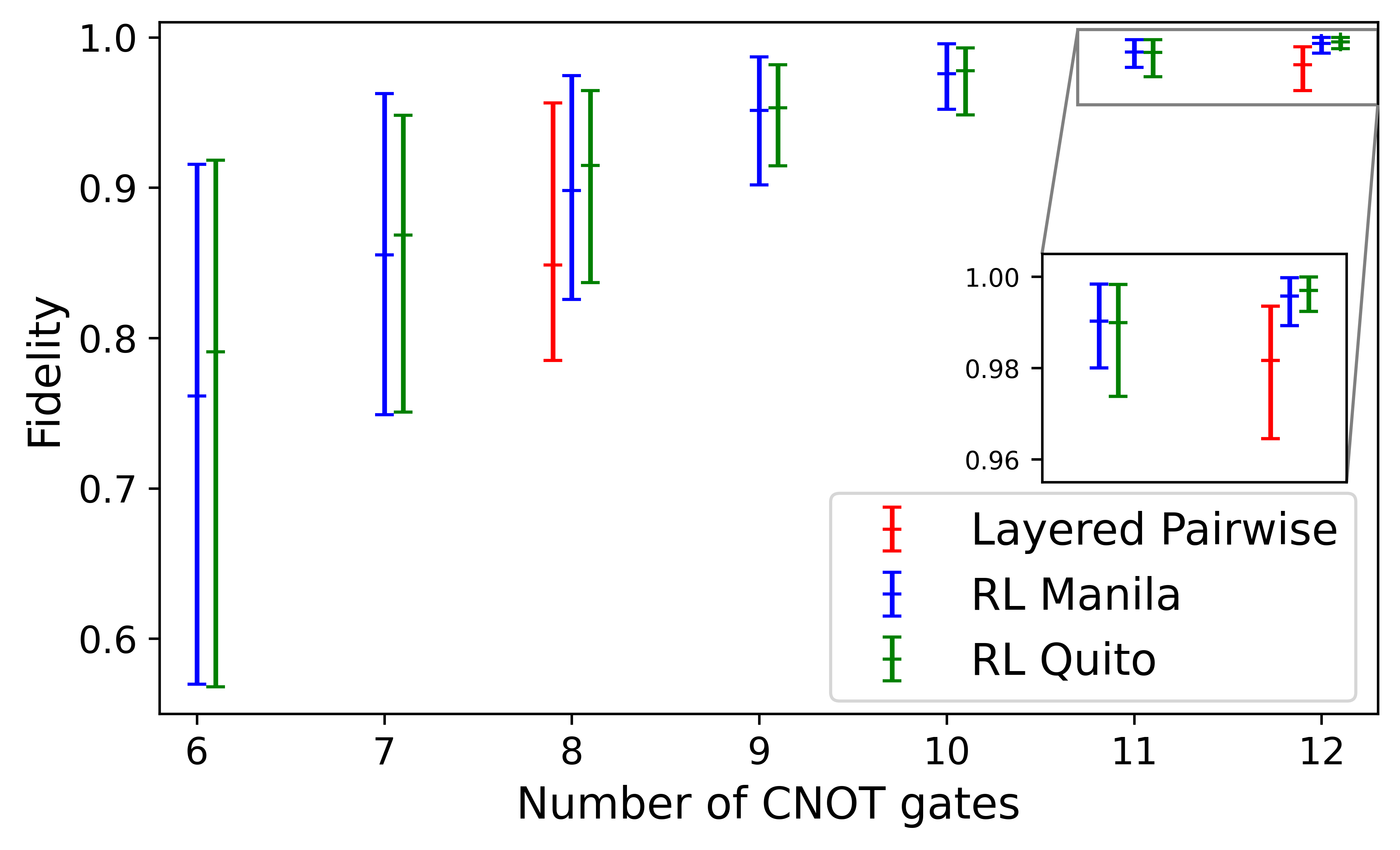}
	\caption{Performance of agents on different CNOT connectivities and of the pairwise hardware-efficient ansatz from Figure \ref{FigLayeredExamples} on a simulated five-qubit system without noise. Agents were restricted to fixed CNOT numbers, the exact x-value has been slightly shifted only for the visualization. NN agents have been limited to the systems connectivity. Layered approaches can only add full layers, thus less data points are available. Error bars depict the smallest interval containing 95\% of the 100 total data samples.}
	\label{FigPerfComp}
\end{figure}

\begin{figure}
	\centering
	\includegraphics[width=1\linewidth]{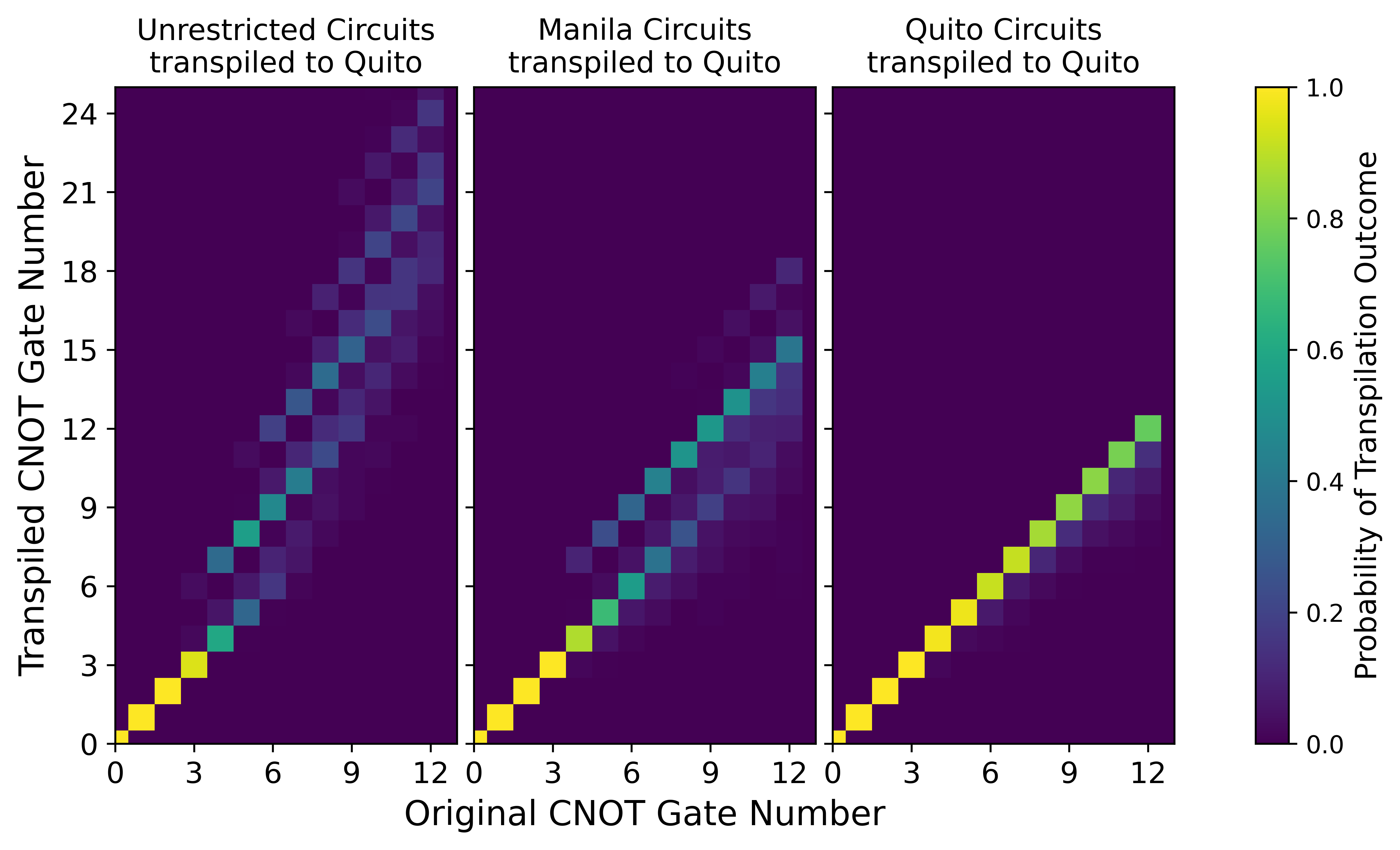}
	\caption{Effects of IBM transpiling with optimization level 3 on random single-qubit and CNOT gate sequences of varying length for five qubits on ibm\_quito. Labeling above the circuit shows the connectivity layout from which the random circuit was generated. If the CNOT connectivity is not taken into account the CNOT gate number rises noticeably. At low gate counts, qubit relabeling can recover circuits that do not require gate replacement. }
	\label{FigTranspiling}
\end{figure}

Figure \ref{FigPerfComp} compares agents against a hardware-efficient ansatz for an ideal system using the architecture of the ibm\_manila and ibm\_quito processors (see Figure \ref{CNOTConnectivity}). For the tested connectivities we are able to outperform the two layered circuit designs in Figure \ref{FigLayeredExamples} at any number of layers, though we only show results for the pairwise architecture as it performed marginally better.
Moreover, we can limit the maximally used number of CNOT gates by forcing the episodes to stop earlier in training and deployment. This may be especially beneficial for NISQ systems, as one can find a trade-off between the approximation error of a desired state and the noise introduced by adding more gates.
The two depicted NN agents perform very similarly despite the difference in CNOT connectivity between ibm\_manila and ibm\_quito, demonstrating the adaptability of the approach. Figure \ref{FigTranspiling} shows the importance of matching the experimental hardware. 
Even in the case that the circuit and hardware architecture differ in only one CNOT connection, one typically expects three extra CNOT gates to be added to circuits when using IBM's built in transpiler. 
While more advanced transpilation and circuit optimization 
techniques exist, these can be computationally expensive or work best for non-universal gate sets (e.g. Clifford) or sparse states. Thus they can struggle with arbitrary states, which we consider. \par

\begin{figure}
	\centering
	\includegraphics[width=1\linewidth]{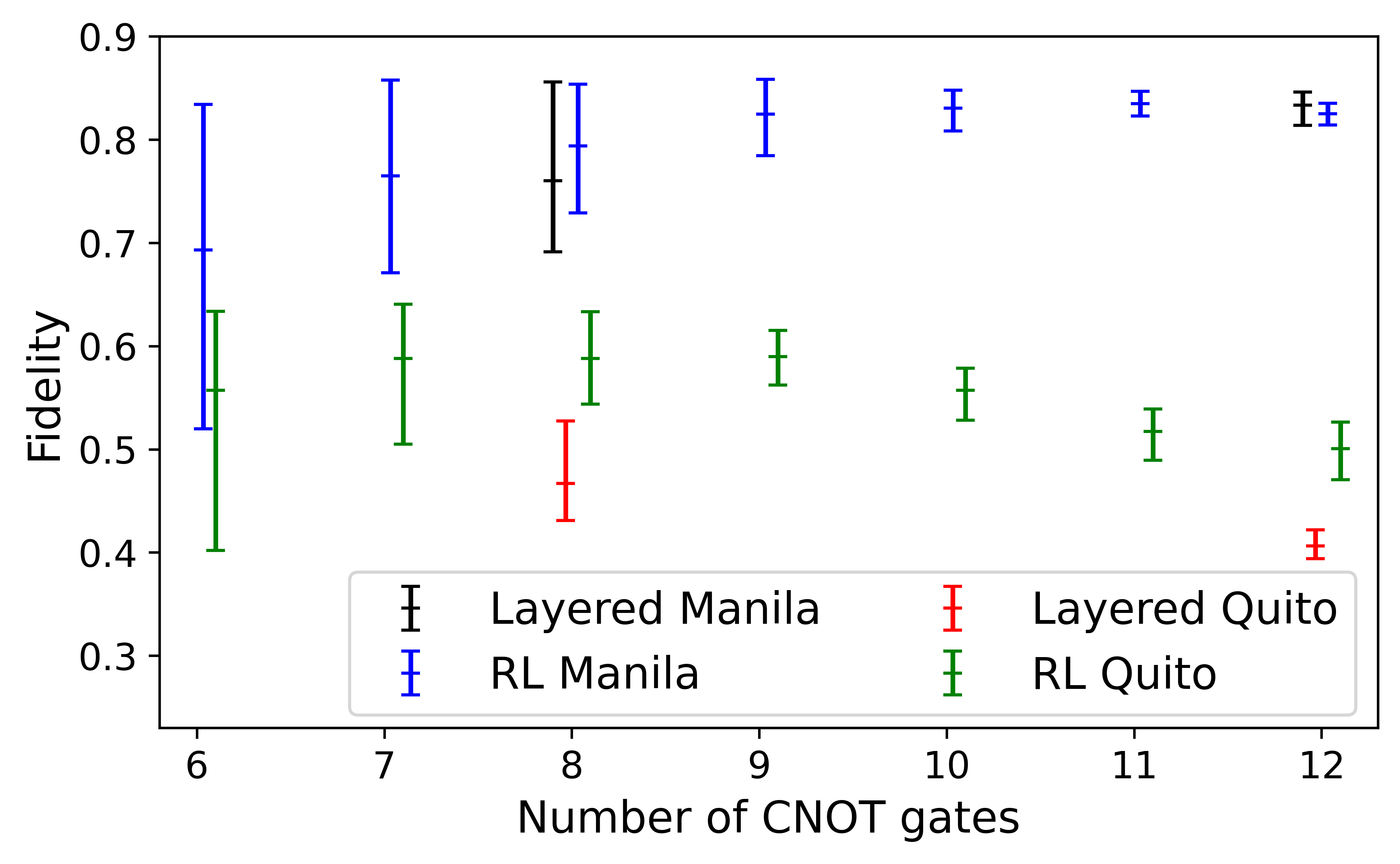}
	\caption{Performance of agents on noisy simulations of IBM backends with readout error disabled. Visualization analogous to Fig. \ref{FigPerfComp}. One can clearly see that using the RL agents we reach optimal performance below 12 CNOT gates. On ibm\_manila the RL agent and layered circuit perform quite similarly though our approach allows for arbitrary CNOT number. For ibm\_quito, which is more impacted by noise in general, there is also a drastic difference between the RL agent and layered circuit due to transpiling effects.}
	\label{FigSimResults}
\end{figure}

In order to verify that an actual advantage on experimental systems should be achievable we have 
tested our circuits on simulations of IBM quantum systems. The fidelities reached on ibm\_manila and ibm\_quito
are depicted in Figure \ref{FigSimResults}. 

We have used the respective fake backends accessible through qiskit, which provide a past snapshot with noise parameters measured on the actual device. To reconstruct the density matrix $\rho$ and calculate the fidelity to the target state, we perform a full quantum state tomography with maximum 
likelihood estimation \cite{Hradil2004}. Readout errors were discarded from the simulation, 
because our focus lies only on improving the circuit gate sequence, however this should not affect our overall conclusions. On an actual physical device one has to use readout error mitigation techniques instead \cite{Cai2023}. \par 
The most immediate observation is that significantly better fidelities are achieved on ibm\_manila in general. This is due to lower gate errors and longer dephasing and relaxation times. Though it should be noted that for ibm\_quito we typically observed higher fidelities on the physical device than in simulations.
This can be seen in Table \ref{tab:SimRealComparison} comparing previous runs between the real and simulated versions for ibm\_quito with only local optimization. When simulating ibm\_quito and using the global optimizer (see Figure \ref{FigSimResults}), we can observe a maximum in the mean fidelity attained by the RL agent at 9 CNOT gates. For ibm\_manila the fidelity peaks at 11 CNOT. This difference can also be understood as a result of noise: As each gate incurs a bigger penalty with increasing noise, the optimum between the approximation error and circuit noise shifts in favour of higher allowable approximation errors. Performance of the layered circuit differs markedly between the experimental devices. While on manila it is close to the RL agent, on quito the necessity of inserting additional CNOT gates during transpiling results in a significant drop in fidelity.

\begin{table}
    \centering
    \begin{tabular}{|c|c|c|c|} \hline
    Type & real & sim & sim\\ \hline
    Optimizer & local & local & global \\ \hline
    Mean Fidelity & $0.693$ & $0.56$ & $0.594$ \\ \hline
    Interval & $\pm0.053$ & $\pm0.12$ & $\pm0.019$ \\\hline
    \end{tabular}
    \caption{Comparison of performance on real device and in simulation for ibm\_quito, including measurement noise. Means and error bars are obtained over 78 four-qubit Haar-random target states. For ibm\_quito the real device results are much better than the simulation. We assume that the difference is caused by the noise parameters of the device snapshot used for the simulator. 
    One can see that the global optimizer is noticeably better than the local optimizer for simulated results. A similar improvement should be expected for real device results.}
    \label{tab:SimRealComparison}
\end{table}

\begin{figure}
	\centering
	\includegraphics[width=1\linewidth]{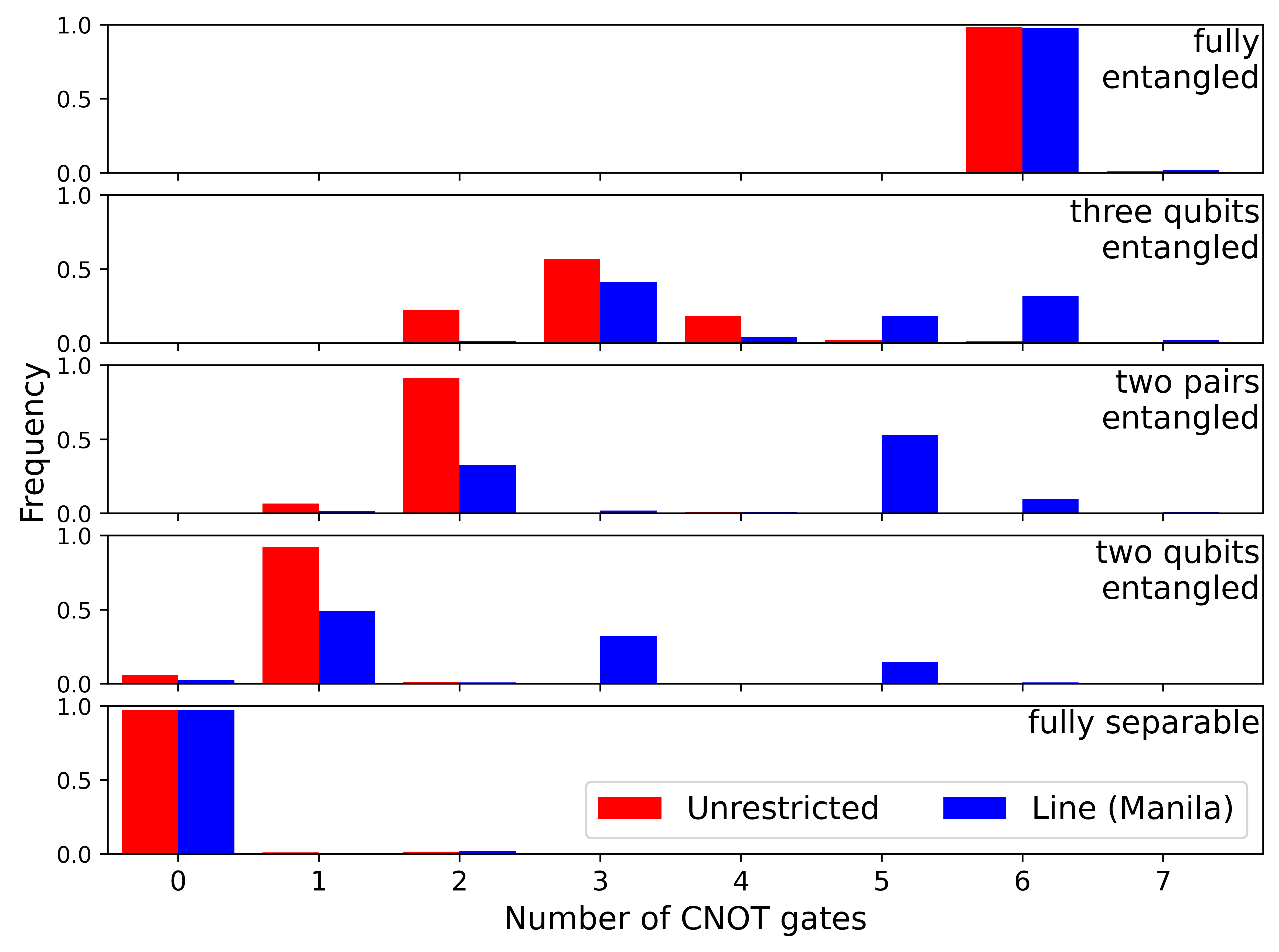}
	\caption{Frequency of the number of applied CNOT gates over 1000 states for each entanglement structure. The "Unrestricted" data points are from an RL agent with all CNOT connections available, the "Line" agent only has connections to the nearest qubits by index. To create states for each separable subsystem Haar random unitaries are sampled. 
		Qubit positions were shuffled. Depicted agents generate circuits with above $0.999$ mean fidelity to the target. The Line agent requires more gates if the entangled qubits are separated with respect to its connectivity.}
	\label{FigCNOTFrequency}
\end{figure}

We have also investigated the performance of our trained agents for separable states on four-qubit systems (see Fig. \ref{FigCNOTFrequency}). Here, results of agents with restricted connectivities may differ significantly from those of agents with full qubit connectivity. For example, a two-qubit entangled state can only be reached using a single CNOT if the direct connection is part 
of the gate set. For short circuits a relabeling step can circumvent this problem. 

\begin{figure}
    \centering
    \begin{tikzpicture}
    \node[scale=0.65] {
    \begin{quantikz}[row sep={0.1cm}]
     &\gate{R_Y}&\ctrl{1}&          &          &          &\gate{R_Y}&\ctrl{1}  &          &        &\gate{R_Y}& \\
     &\gate{R_Y}&\targ{} &\gate{R_Y}&\ctrl{1}  &          &\gate{R_Y}&\targ{}   &\gate{R_Y}&\ctrl{1}&\gate{R_Y}&\\
     &          &        &\gate{R_Y}&\targ{}   &\gate{R_Y}&\ctrl{1}  &          &\gate{R_Y}&\targ{} &\gate{R_Y}& \\
     &          &        &          &          &\gate{R_Y}&\targ{}   &          &          &        &\gate{R_Y}& \\
    \end{quantikz}};
    \end{tikzpicture}%
    \caption{Circuit found by the RL agent for the four-qubit W-state that may be generalized to $n$ qubits.
    }
    \label{fig:CircWstate}
\end{figure}
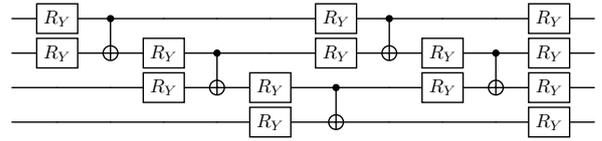

In order to guarantee that agents can learn to adapt to individual targets, we have checked their performance on well-known fully entangled states with proven optimal implementations. Specifically, this was done for the GHZ-state and W-state. The four-qubit agents correctly identify circuits with a lower number of entangling gates when compared to a random target, demonstrating that the agent is not just using highly expressive circuits. This is remarkable, 
as the GHZ-state and W-state were almost certainly never observed directly in the training data. 
On 10 datasets containing 2 million Haar random states each, we observed a maximum fidelity to a W- or GHZ-state of around $0.678$.
We estimate a probability of less than $0.1\%$ that the highest fidelity to the W-state or GHZ-state in a training set was above $0.75$. 
For GHZ states the number of employed entangling gates is equal to the known optimal gate number of $n-1$, i.e. three CNOTs on four qubits. For W-states, accurate circuits containing only five CNOT were found. The general structure for the four-qubit W-state circuit (see Figure \ref{fig:CircWstate}) can be generalized to arbitrary $n$ and requires only $2(n-1)-1$ CNOTs. We have tested for up to ten qubits that these circuits accurately create the W-state. However, a general expression for the rotation gate angles eludes us.\par 
Agents with thresholds allowing for lower approximation accuracy adjust the gate number to match the target requirements. Circuits remained optimal when restricting the available resources in this way, which we have verified through brute-forcing (see Table \ref{tab:WstatePerf}).

\begin{table}
    \centering
    \begin{tabular}{c|c|l}
      \multicolumn{1}{p{1.6cm}|}{\centering Number \\ of CNOT} & \multicolumn{1}{|p{1.6cm}|}{\centering Optimal \\ Fidelity} & \multicolumn{1}{|p{2.2cm}}{RL Example Sequence} \\ \hline
       3  & 0.8 &  2-3, 1-2, 0-2\\
       4 & 0.933 & 1-2, 2-3, 1-0, 1-2 \\
       5 & 1 & 0-1, 1-2, 2-3, 2-1, 1-0
    \end{tabular}
    \caption{Performance on four-qubit W-states with example circuits extracted from RL. The listed sequence written as a list of gates with each gate in the format "control-target".}
    \label{tab:WstatePerf}
\end{table}

\section{Conclusion}
We have demonstrated an algorithm that is capable of finding quantum gate sequences for arbitrary target states, generating them with a higher fidelity than layered parameterized quantum circuits. 
Tuning the CNOT gate number or threshold allows us to set the approximation accuracy and balance it with respect to experimental noise. For five qubits it was consistently possible to train to above $0.99$ mean fidelity to random target states in simulations.
We expect that, experimentally, our approach will allow for higher state preparation fidelity, especially on systems dominated by entangling gate noise. Combining our algorithm with other circuit optimization protocols could yield even better results. \par
The biggest advantage of our approach is its flexibility and tunability. RL promises to work regardless of the gate set or qubit connectivity. Furthermore, one can choose desired target thresholds for the training, giving the ability to find a trade-off between minimizing experimental noise and approximation error. 
\par
Compared to previous RL methods, which were mostly designed for gate sets without continuous 
parameters, such as the Clifford set or the Toffoli+T gate set, we expect an improvement in the direct applicability to experimental devices with low qubit numbers. The Clifford gate set is not universal, and discrete gate sets are likely to be inefficient in terms of gate number or depth, because for operations which are easily achieved by few continuous parameters they necessitate long chains of discrete gates. However, a potential drawback of our method is the limitation to smaller qubit numbers than non-universal approaches, as the proposed algorithms requires information of the quantum state as input and needs to simulate the state evolution to gather training data. With future development, one may be able to deal with these issues by using an accurate but unoptimized gate sequence as input and by obtaining training data from real NISQ quantum computers. The latter would also immediately give access to the exact noise behaviour. \par 

\section{Acknowledgments}
We acknowledge the use of IBM Quantum services for this work. The views expressed are those of the authors, and do not reflect the official policy or position of IBM or the IBM Quantum team. 
This work was partially finanically supported by the Deutsche Forschungsgemeinschaft
via the International Research Training Group 2676 'Imaging quantum systems (IQS): photons. molecules, materials', grant no. 437567992.

\bibliography{Literature}

\end{document}